\documentclass[twocolumn,showpacs,preprintnumbers,amsmath,amssymb,letterpaper]{revtex4}
\usepackage{graphicx}
\usepackage{dcolumn}
\usepackage{bm}
\newcommand\Eq[1]{Eq.~\ref{eq:#1}}

\newcommand\Tab[1]{Table~\ref{tab:#1}}
\usepackage{amssymb}
\usepackage{epsfig}
\newcommand{\be}{\begin{equation}}
\newcommand{\ee}{\end{equation}}
\newcommand\beq{\begin{eqnarray}}
\newcommand\eeq{\end{eqnarray}} 
\newcommand\eqn[1]{\label{eq:#1}} 
\newcommand\eq[1]{eq. (\ref{eq:#1})}

\newcommand{\vev}[1]{\langle #1 \rangle}

\newcommand{\CA}{{\cal A}}

\newcommand{\CE}{{\cal E}}

\newcommand{\CN}{{\cal N}}

\newcommand\half{{\textstyle{\frac{1}{2}}}} 
 
\newcommand\expect[3]{\langle #1|#2|#3\rangle}

\newcommand{\mybar}[1]%
        {\kern 0.6pt\overline{\kern -0.6pt#1\kern -0.6pt}\kern 0.6pt}
\begin{document}

\preprint{INT-PUB-11-25}
\preprint{RIKEN-QHP-2}

\title{Noise, sign problems, and statistics}

\author{Michael G. Endres$^1$}
\email{endres@riken.jp}

\author{David B. Kaplan$^2$}
\email{dbkaplan@uw.edu}

\author{Jong-Wan Lee$^2$}
 \email{jwlee823@u.washington.edu}

\author{Amy N. Nicholson$^2$}
 \email{amynn@u.washington.edu}

\affiliation{$^1$Theoretical Research Division, RIKEN Nishina Center, Wako, Saitama 351-0198, Japan}

\affiliation{$^2$Institute for Nuclear Theory, University of Washington, Seattle, WA 98195-1550, USA}

 \date{\today}
 
 \begin{abstract}
We show how sign problems in simulations of many-body systems can manifest themselves in the form of heavy-tailed correlator distributions, similar to what is seen in electron propagation through disordered media.  We propose an alternative statistical approach for extracting ground state energies in such systems, illustrating the method with a toy model and with lattice data for unitary fermions.

%

\end{abstract}
\maketitle

\section{Introduction}

 One of the most challenging and interesting problems in physics is to understand the properties of a system of many strongly interacting fermions.  Numerical simulation is an important tool for understanding the ground state, and the common approach is to compute the $N$-body correlator $C_N(\tau;\phi) = \expect{0}{\Psi_N(\tau)\Psi_N^\dagger(0)}{0}_\phi$, where $\Psi_N^\dagger(0)$, $\Psi_N(\tau)$ are interpolating fields which create an $N$-body state at Euclidean time zero and annihilate it at time $\tau$, and $\phi$ is a stochastic field responsible for fermion interactions.  The field $\phi$ could be the dynamical gluon field  in the case of QCD, for example, or an auxiliary field to induce short-range interactions.  For large $\tau$ the averaged  correlator  asymptotically approaches 
 \beq
\langle C_N(\tau,\phi)\rangle \sim Z e^{-\tau E_0(N)}
\eqn{cdef}\eeq
where $E_0(N)$ is the ground state energy of the system and $\sqrt{Z}$ is the amplitude for $\Psi$ to create the ground state.   Therefore if one computes
$ -\frac{1}{\tau}\ln \mybar C_N(\tau)
 $, where $\mybar C_N(\tau) =\frac{1}{{\cal N}} \sum_i C_N(\tau,\phi_i) $ is a sample mean computed on an ensemble of $\CN$ statistically independent $\phi$ fields, one expects to see a ``plateau" at large $\tau$ whose height yields the ground state energy $E_0(N)$.   Excited state energies and response of the ground state to probes can also be computed by variations of this technique.
 
The computation of  $-\frac{1}{\tau}\ln \mybar C_N(\tau)$ can be problematic, however:  it might be excessively noisy, or it may drift with $\tau$ and never find a plateau.  We wish to address these problems here, defining the former as a ``noise" problem, and the latter as an ``overlap" problem, both of which can be related to the sign problem encountered in  lattice  simulations at nonzero chemical potential.  In particular, referring to recent lattice simulations by the present authors of large numbers of unitary fermions,  we show that the problems encountered can be manifestations of heavy-tailed distributions for $C_N(\tau,\phi)$ which make computing $ \ln \langle C_N\rangle$ very difficult, and that the ideal estimator for this quantity might not simply be $\ln \mybar C_N$, as is commonly used.  We find here that a cumulant expansion in the log of the correlator is a more efficient estimator, for example.  More generally, we suggest that a study of the statistics of systems exhibiting noise or an overlap problem might be exploited to greatly facilitate the extraction of useful physics from numerical simulations.

\section{Noise, and the physical spectrum}

\begin{figure*}
\begin{tabular}{cc}
\includegraphics[width=7 cm]{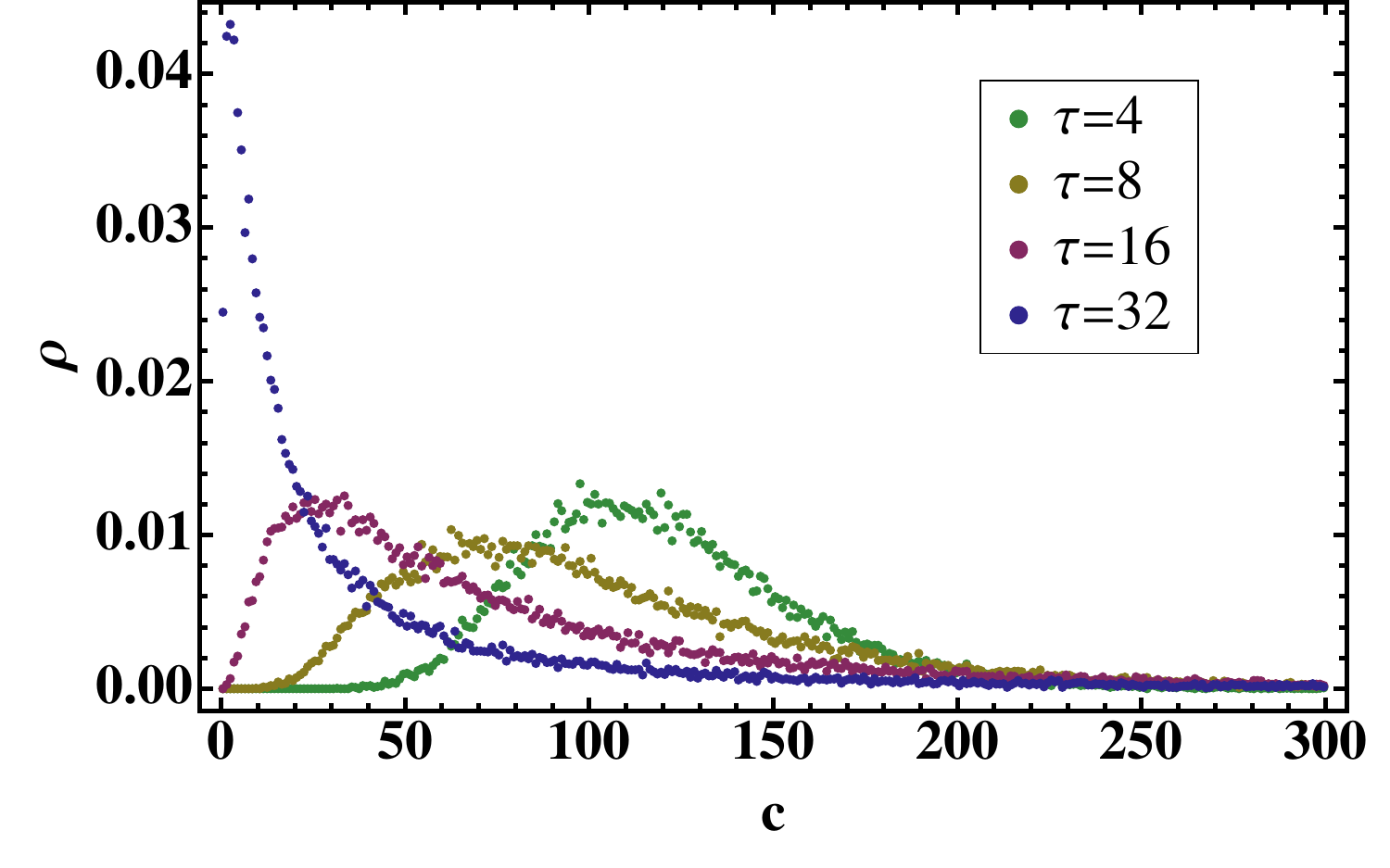}
\includegraphics[width=7 cm]{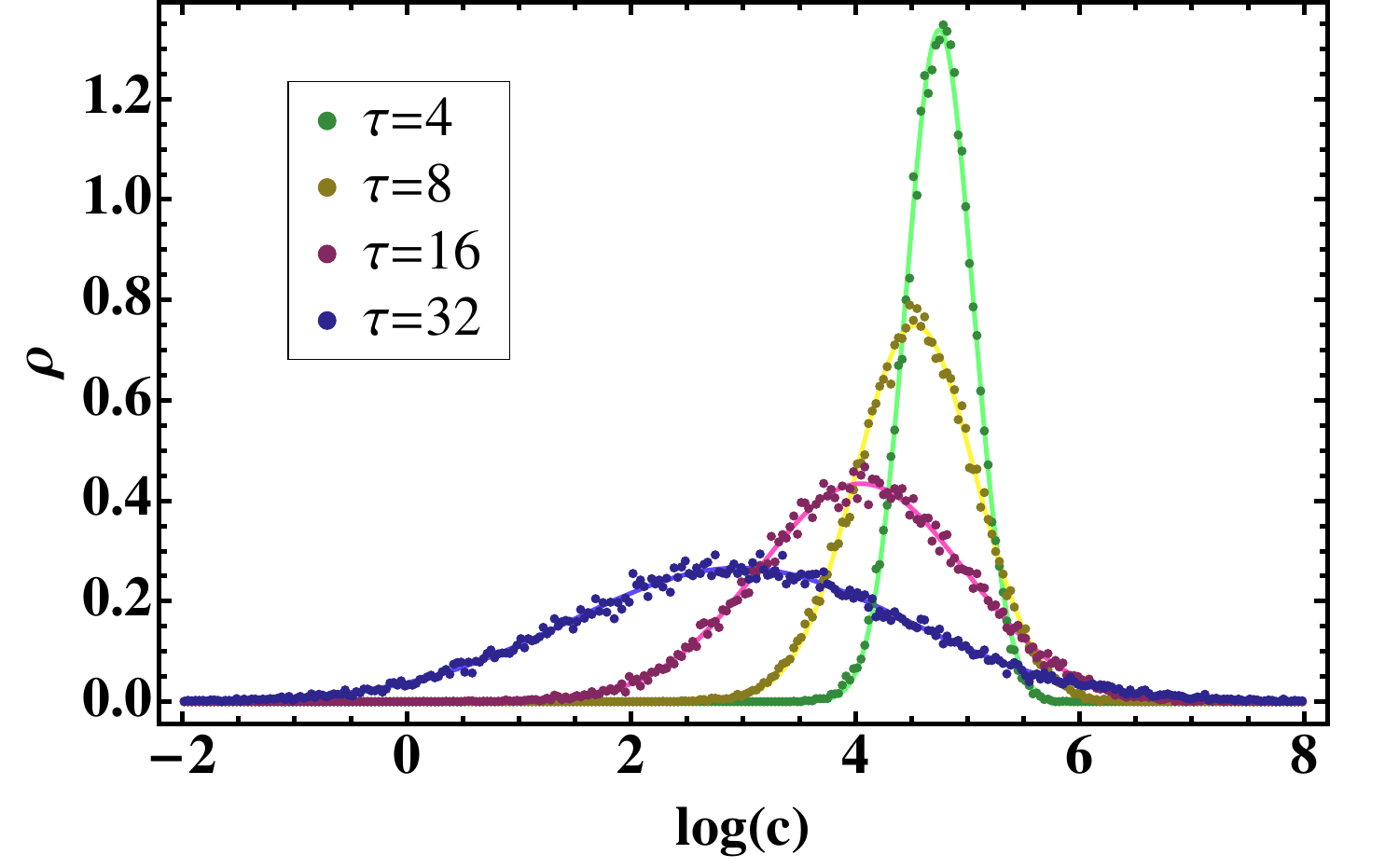}
\end{tabular}%
\caption{\label{fig:ln} {Histograms for distributions of $c=C_N(\tau\phi)$ and $\ln (c)$ for $N=4$ unitary fermions at several times $\tau$, taken from Ref.~\cite{unitary:2011aa}.  Curves fitting $\ln(c)$ are Gaussian, implying that $c$ is approximately log-normal distributed, with $\sigma^2$ increasing with time.  } }
\end{figure*}

 The sign problem encountered in $N$-particle simulations does not arise simply because of Fermi statistics; if that were the only obstacle one could construct $C_N$ as an $N\times N$ Slater determinant of one-body propagators,  with a computational difficulty of computing the determinant scaling only as $N^3$.  In contrast, the sign problems commonly encountered, such as with QCD at nonzero chemical potential, entail computational difficulty which grows exponentially with particle number; furthermore, serious sign problems can occur in bosonic systems as well. Instead, sign problems appear when there are multiparticle states for which the energy/constituent is lower than for the states one wants to study.    For example, if $\langle C_A\rangle\sim e^{- M_A \tau}$ is the expectation of a $3A$ quark correlator in QCD for a nucleus  of atomic number $A$ and mass $M_A$, the  variance in the sample mean $\mybar C_A$  can be estimated as 
\beq
\sigma^2\sim \langle C_A^\dagger C_A\rangle -\langle C_A^\dagger\rangle\langle C_A\rangle \sim \frac{1}{\CN} e^{-3A m_\pi \tau}
\eeq 
 for sample size $\CN$.  Since $C_A$ corresponds to $3A$ quark propagators and $C_A^\dagger$ to $3A$ anti-quark propagators, the variance is dominated by the state with $3A$ pions and $\sigma$ falls off with $\tau$ much more slowly than the signal one is looking for, $\langle C_A\rangle$, since $\frac{3}{2} m_\pi A\ll M_A$.  This  ``Lepage analysis" \cite{Lepage:1989hd}  suggests there is a noise problem and that it arises because in a background gluon field each quark propagator is uncorrelated with any other and doesn't ``know" whether it is to be contained in a light pion or a heavy nucleon.  This suggests a picture where each correlator $C_A(\tau,\CA)$ in a particular background gauge field $\CA$  roughly equals $e^{-3A/2 m_\pi \tau}$, and the exponentially smaller value expected for $\langle C_A\rangle$ only arises from subsequent cancellations while averaging over gauge fields.  A very similar analysis applies to QCD with nonzero chemical potential \cite{Gibbs:1986xg,Splittorff:2006fu}.
    This would be a reasonable picture if the distribution of $C_A(\tau,\CA)$ over the ensemble of gauge fields was normal, with mean $ e^{- M_A \tau}$ and variance $ e^{-3A m_\pi \tau}$, with  large fluctuations concealing an exponentially small signal.  There are general arguments that suggest this is incorrect, however, and that the distribution of many-fermion correlation functions will be heavy-tailed and extremely non-Gaussian, a result we also find from explicit simulations of unitary fermions.  In the latter case we show that better understanding the nature of the noise can help devise an efficient strategy for extracting a signal; it is plausible that similar techniques could be more widely applicable to noisy systems.

\section{A Mean Field Description} 
 
Nonrelativistic fermions with strong short-range interactions tuned to a conformal fixed point where the phase shift satisfies $\delta(k)=\pi/2$ for all $k$  are called ``unitary fermions".   This nonrelativistic conformal field theory is interesting to study both for its simplicity and universality, its challenges for many-body theory, and because it can be realized and studied experimentally using trapped atoms tuned to a Feshbach resonance. It is also an ideal theory for studying fermion sign problems on the lattice, being much simpler and faster to simulate than QCD.   At its most basic, the lattice action is the obvious discretization of the Euclidean Lagrangian \cite{ Chen:2003vy}
\beq
\psi^\dagger (\partial_\tau -\nabla^2/2M)\psi -\half m^2 \phi^2 + \phi \psi^\dagger\psi
\eqn{lag}
\eeq
 where $\phi$ is a nonpropagating auxiliary field with $m^2$  tuned to a critical value $m^2_c$, and $\psi$ is a spin $\half$ fermion with mass $M$; a more sophisticated action tuned to reduce discretization errors was recently presented in \cite{Endres:2010sq}.
 A simulation of this theory   reveals a distribution for $N$-body correlators $C_N(\tau,\phi)$ which is increasingly non-Gaussian at late $\tau$; in fact, it is $\ln C_N$ which appears to be roughly normally distributed, as shown in Fig.~\ref{fig:ln}, so that  $C_N(\tau,\phi)$  is roughly log-normal distributed with an increasingly large $\sigma$ and long tail at late time.

\begin{figure}[b]
\includegraphics[width=7 cm]{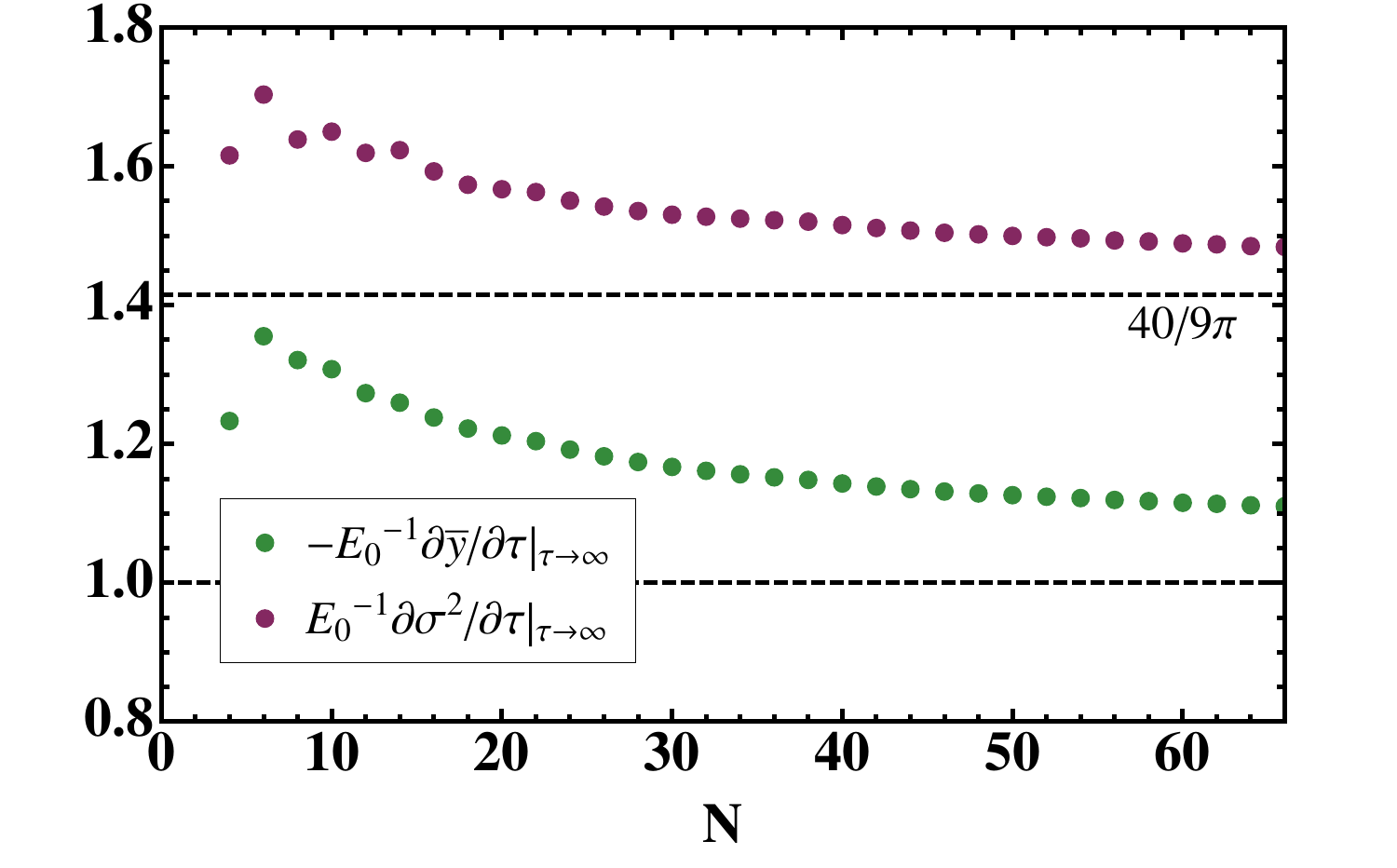}
\caption{\label{fig:sigmu} {The quantities $-\frac{1}{E_0}\frac{\partial\mybar y}{\partial\tau}$ and $\frac{1}{E_0}\frac{\partial\sigma^2}{\partial\tau}$ as a function of $N$ for unitary fermions at late times on a lattice of size $L=10$, compared to mean field prediction \eq{mf} (dashed lines).}}
\end{figure}

The appearance of a heavy-tailed distribution should not be surprising, since the system is similar to the problem of electron propagation in disordered media, where  heavy-tailed distributions are ubiquitous  in the vicinity of  the Anderson localization transition. For example, it is found that for physical quantities such as the current relaxation time or normalized local density of states, the distribution function $P(z)$ scales as $\exp(-C_d \ln^dz)$. A particularly simple way to derive these results is to use the optimal fluctuation method of Ref.~\cite{smolyarenko1997statistics}, which is a mean field approach.  We can adapt these methods to the current problem, defining the variable $Y=\ln C_N(\tau,\phi)$ and computing its probability distribution $P(y)$ as
\beq
P(y) &=& {\cal N} \int D\phi e^{-S_\phi}\, \delta(Y(\tau,\phi)-y) \cr
     &=& {\cal N} \int D\phi\,\frac{dt}{2\pi} e^{-S}
\eeq
where $S_\phi = \int d^4x \textstyle{\frac{m^2}{2}}\phi^2$ and  $S= S_\phi -it(\ln C_N(\tau,\phi)-y)$. Using the PDS subtraction scheme \cite{Kaplan:1998tg}
we have $m^2 = M\lambda/4\pi$, where the renormalization scale $\lambda$ is taken to be the physical momentum scale in the problem --- in this case  $\lambda = k_F \equiv (3\pi^2N/V)^{1/3}$, $N/2$ being the number of fermions with a single spin orientation. We proceed now to evaluate this integral using a mean field expansion; it is not evident that there is a small parameter to justify this expansion, but the leading order  result is illuminating and fits the  numerical data well.  We expand about $\phi(x)=\phi_0$, $t=t_0$, and use the fact that  for large $\tau$ the $n^{th}$ functional derivative of $\ln C_N(\tau,\phi)$ with respect to $\phi(x)$  equals the the  1-loop Feynman diagram with $n$ insertions of $\psi^\dagger\psi$ in the presence of a chemical potential $\mu=k_F^2/(2 M)$.  The equations for $\phi_0$ and $t_0$ are given by
\beq
t_0 &=& -i \frac{m^2\phi_0}{\vev{n(x)}_c}=-i\frac{V m^2\phi_0}{N}\cr
\phi_0 &=&- \frac{y-\ln Z + \tau E_0(N)}{N\tau}
\eeq
where $E_0(N) = 3N E_F/5 $ is the total energy of $N$ free degenerate fermions ($N/2$ of each spin), and $Z$ is the overlap of the source and sink with the free fermion state. The leading term in the mean field expansion for $P(y)$ can therefore be expressed as
$P(y)\propto \exp\left[-\frac{(y-\mybar y)^2}{2\sigma^2}\right]
$
with
\beq
\mybar y = \ln Z -\tau E_0(N)\ ,\quad \sigma^2 = \frac{40}{9\pi} E_0(N)\,\tau\ .
\eqn{mf}
\eeq
This describes a log-normal distribution for the $N$-fermion propagator $C_N(\tau,\phi)$, with both mean and variance growing with time in units of the energy of $N$ free degenerate fermions. In Fig.~\ref{fig:sigmu} we plot the quantities $-\frac{1}{E_0}\frac{\partial\mybar y}{\partial\tau}$ and $\frac{1}{E_0}\frac{\partial\sigma^2}{\partial\tau}$ as a function of $N$ obtained from correlator distribution data for unitary fermions at late $\tau$, and find that the gross features of the results are compatible with the mean field estimates of unity and $40/9\pi$ obtained from \eq{mf}.

 \section{A toy model}

It would be useful to devise an algorithm to reliably estimate energies without having to exhaustively sample the long tail of the correlator distribution, yet without making incorrect assumptions about the exact functional form of that tail.  An approach we suggest here is to exploit the general relationship between stochastic variables $X$ and $Y=\ln X$:
\beq
\ln\vev{X} = \sum_{n=1}^\infty\frac{\kappa_n}{n!}
\eqn{alg}
\eeq
where $\kappa_n$ is the $n^{th}$ cumulant of $Y$.  This relation can be proved by noting that the generating function for the $\kappa_n$ is $\ln \phi_Y(t)$ where $\phi_Y(t)=\vev{e^{Y t}}= \vev{X^t}$ is the moment generating function for $Y$, and evaluating at $t=1$,  assumed to be  within the radius of convergence.  The motivation for investigating \eq{alg} is that if the distribution $P(X)$ were exactly log-normal, the above sum would end after the second term, as $\kappa_{n>2}$ would all vanish; therefore by replacing the $\kappa_n$ by sampled cumulants and truncating the sum at finite order, one might hope to have a reliable estimator for $ \ln\vev{X} $ provided $P(X)$ was nearly log-normal, in the sense that the $\kappa_n$ fall off rapidly for $n>2$.

Distributions with log-normal-like tails arise naturally in products of stochastic variables.  The propagator $C_N(\tau,\phi)$ for unitary fermions can be expressed in a transfer matrix formalism as the  product of a $\tau$ matrices --- one per time hop --- each of which is the direct product of $N$ $V\times V$ matrices  of the form $e^{-K/2} (1+ g  \varphi) e^{-K/2}$,  where $K$ is a constant matrix (the spatial kinetic operator), $ \varphi$ is a random  diagonal matrix with $O(1)$ entries corresponding to stochastic $\phi$ fields living on the time links, and $g$ is a coupling constant (identified with $1/m^2$ in \Eq{lag}) that has been tuned to a particular critical value that is $O(1)$.  Unfortunately, little seems to be known about products of random matrices, beyond the study in \cite{Jackson:2002qx} which deals with large products of weakly random matrices.  Therefore we analyze instead a toy model where we define a ``correlator" $C_\tau$  as a product of random numbers, and an ``energy" $\CE = \lim_{\tau\to\infty} \CE_\tau$  where:
\beq
C_\tau= \prod_{i=1}^\tau (1+ g \varphi_i)\ ,\quad \CE_\tau= -\frac{1}{\tau} \ln\vev{C_\tau}
\eqn{model}\eeq
where $0\le g\le 1$ and the $ \varphi_i$ are independent and identically distributed random numbers with a uniform distribution on the interval $[-1,1]$.  The exact value for the energy is obviously $\CE_\tau=0$ for any $\tau$ since the statistical average of the correlator is $\vev{C_\tau}=1$. 
The cumulants of the variable $Y=\ln (C_\tau)$ are given by
\beq
\kappa_1  &=&\tau \left [\textstyle{ \frac{1}{2}} \log \left(1-g^2\right)+\textstyle{\frac{\tanh ^{-1}(g)}{g}}-1\right]\ ,\cr
\frac{\kappa_{n}}{n!} &=&\tau\left(\textstyle{ \frac{(-1)^n}{n}}- \text{Li}_{1-n}\left(\textstyle{\frac{1+g}{1-g}}\right)\textstyle{ \frac{\left(2\tanh ^{-1}(g)\right)^n}{n!}}\right)\nonumber
\eeq
for $n\ge 2$; for small $g$ one finds that the $\kappa_n $ rapidly decrease as $n$ increases for $g<1$. Table~\ref{tab:table1} shows how the systematic error in \eq{alg} when truncated at $n=n_{max}$, converges to the exact answer $\CE_\tau=0$ as a function of $n_{\max}$ for $g=1/2$, and shows that even though the distribution is not log-normal ($\kappa_{n>2}\ne0$) the convergence is rapid.

  
\begin{figure}[t]
\includegraphics[width=7.0 cm]{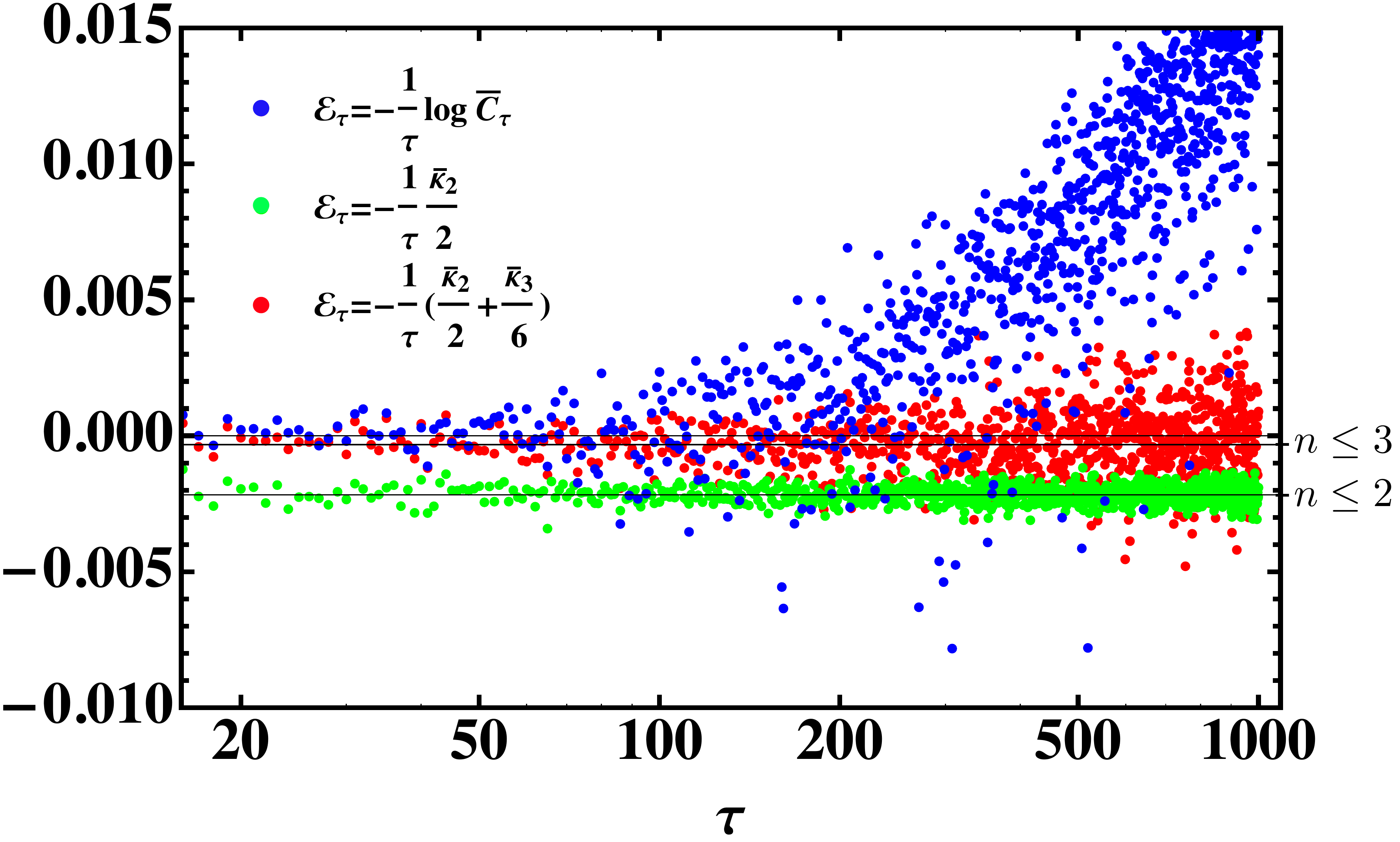}
\caption{Simulation of the energy $\CE_\tau$ for the model \eq{model} with $g=\half$.   The exact answer is $\CE_\tau=0$ (black line);    exact values of \eq{alg} truncated at order $n=2,3$ are indicated.} 
\label{fig:sim}
\end{figure}

\begin{table}[t]
\caption{\label{tab:table1} $\CE$ determined from 250 blocks of 50,000 configurations each for the model \eq{model} with $\tau=1000$, $g=1/2$.  }
\begin{ruledtabular}
\begin{tabular}{crrr}
Method & $\CE\quad\ $ & stat. error & syst. error \\
\hline
conventional & 0.014932 & 0.002485 & -- \\
$\kappa_{n\le 2}$ &-0.002159& 0.000304& -0.002165\\
$\kappa_{n\le 3}$&-0.000412 & 0.001618&-0.000324 \\
$\kappa_{n\le 4}$&-0.000647 & 0.008379&0.000050 \\
$\kappa_{n\le 5}$&-0.001794 & 0.037561&$3.34\times10^{-6}$ \\
$\kappa_{n\le 6}$&0.010943 & 0.147739&$-1.22\times10^{-6}$ \\
 \end{tabular}
\end{ruledtabular}
\end{table}

\begin{figure}
\includegraphics[width=6.8 cm]{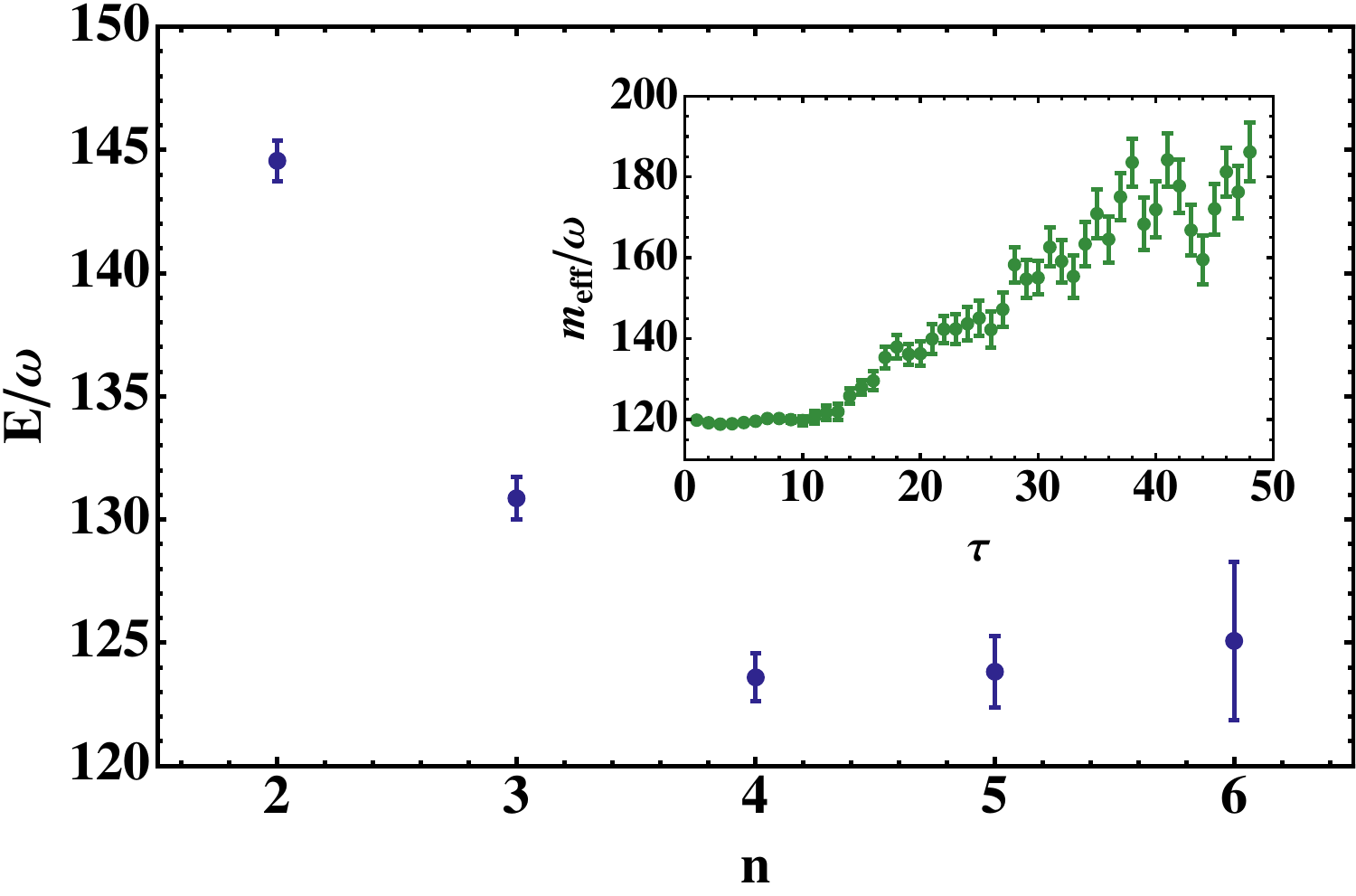}
\caption{ Energy for 50 unitary fermions in a harmonic trap,  $10^6$ configurations; fits performed using expansion \eq{alg} up to order $n$ over the time interval 45-60 (n=2,3) and 13-60 (n=4,5,6).
Inset: conventional effective mass.} 
\label{fig:Convergence}
\end{figure}

 In Fig.~\ref{fig:sim} we show the results of a simulation where we compute $\CE_\tau$ for $g=\half$ and $\tau=1,\ldots,1000$.  At each value of $\tau$ we independently generated an ensemble of values for $C_\tau$ of size $N=50,000$.  From that ensemble we computed $\CE_\tau$ by 
 (i)  using the conventional estimator $\CE_\tau = -\frac{1}{\tau} \ln \mybar C_\tau$ (blue), which shows a striking systematic error for $\tau\gtrsim 50$, and statistical noise increasing up to $\tau\simeq 500$ but decreasing beyond that;
  (ii) using \eq{alg} truncated at $n=2$ using conventional estimators for the $\kappa_n$ (green), showing a $\tau$-independent systematic error with smaller but slowly  growing statistical error; (iii)  \eq{alg} truncated at $n=3$ (red) with a negligible constant systematic error but a larger statistical error, growing with $\tau$.  Evidently, one trades systematic error for statistical error by truncating \eq{alg} at increasingly large $n_{max}$. %
%
%
%

\Tab{table1} displays results of a simulation of  $1.25 \times 10^7$ $\phi$ configurations blocked into 250 blocks of 50,000 each, for the model \eq{model} at $\tau=1000$ and $g=1/2$.  For each case we give the mean and the square root of the variance; for the truncated cumulant expansion we also give the theoretical systematic error from truncating \eq{alg} using our analytic expressions for $\kappa_n$. These numbers show how the conventional method gives a wrong answer with deceptively small statistical error. One sees again the trade of systematic error for statistical error as one increases the order $n_{max}$ where one truncates the sum in \eq{alg}.  \Tab{table1} suggests the place to stop for the smallest combined error is at $n_{max}=3$, justified by noting that the $n_{max}=4$ result with statistical errors encompasses the $n_{max}=3$ result; we suggest this as a practical algorithm for determining where to truncate the cumulant expansion in general.
Fig.~\ref{fig:Convergence}  shows how this works in a real simulation for 50 trapped unitary fermions \cite{unitary:2011aa}.


\section{Discussion}

Heavy-tail distributions  are likely to be ubiquitous in $N$-body simulations, and perhaps even in other types of noisy calculations. With such distributions theoretical statistical means can deviate wildly from sample means for any realizable sample size and render  conventional estimates of expected fluctuations irrelevant.   We have shown that there are more efficient estimators for ground state energies using the cumulants of the log of the correlator instead of the conventional effective mass, at least for positive correlators.  This method  is presumably only effective for nonpositive data when when the heavy-tail is asymmetric.  It may be useful to think of this procedure in a renormalization group language, where the higher cumulants behave like irrelevant operators affecting the flow toward a log-normal distribution.   %
%

\begin{acknowledgments}
This work was supported in part by U.S.\ DOE grant No.\ DE-FG02-00ER41132.
M.G.E is supported by the Foreign Postdoctoral Researcher program at RIKEN.
\end{acknowledgments}
\bibliography{Stats}
\end{document}